\documentstyle[aps,prd,psfig]{revtex}
\begin{document}
\newcommand{\bec}{\begin{center}}
\newcommand{\eec}{\end{center}}
\renewcommand{\thesection}{\arabic{section}}
\renewcommand{\thesubsection}{\arabic{subsection}}

\title{Effect of Strong Quantizing Magnetic Field on the Transport
Properties of Dense Stellar Plasma}

\author{
Soma Mandal$^{a)}$ and Somenath Chakrabarty$^{a),b)}
 ${\thanks{E-Mail: somenath@klyuniv.ernet.in}}}
\address{$^{a)}$Department of Physics, University of Kalyani, Kalyani 741 235,
India and $^{b)}$Inter-University Centre for Astronomy and Astrophysics, Post 
Bag 4, Ganeshkhind, Pune 411 007, India}
\date{\today}
\maketitle
\begin{abstract}
The transport properties of dense stellar electron-proton plasma is
studied following an exact relativistic formalism in presence of strong 
quantizing magnetic field. The variation of transport coefficients with
magnetic fields are found to be insensitive for the field strengths 
$\leq 10^{17}$G, beyond which all of them abruptly go to zero. As a 
consequence, the electron-proton plasma behaves like a superfluid insulator 
in presence of ultra-strong magnetic field.  
\end{abstract}
\section{Introduction}
The study of the effect of strong magnetic field on dense stellar plasma
is one of the oldest branches of physics. It has gotten a new life
after the discovery of a few magnetars- the strange stellar objects,
with unusually high surface magnetic fields \cite{R1,R2,R3,R4}. 
These stellar objects are believed to be strongly magnetized young neutron 
stars. The surface magnetic fields are observed to be $\sim 10^{15}$G. Then 
it is quite possible that the field strength at the core region may go up to 
$10^{18}$G.  The exact source of this strong magnetic field is of course yet 
to be known. These objects are also supposed to be the possible sources of 
anomalous X-ray and soft gamma emissions (AXP and SGR). If the magnetic fields 
are really so strong, in particular at the core region, they must affect 
significantly most of the important physical properties of such stellar 
objects and the physical processes, e.g., weak and electromagnetic interactions
taking place at the core region. Which means, the presence of strong quantizing
magnetic field at the core region should modify, both qualitatively and 
quantitatively the equation of state of dense neutron star matter,
and as a consequence the gross-properties of neutron stars 
\cite{R5,R6,R7,R8}, e.g., mass-radius relation, moment of inertia,
rotational frequency etc. should also change significantly. In the case of
compact neutron stars, the phase transition from neutron matter to quark 
matter at the core region, if any, will also be affected by strong quantizing 
magnetic field.  It has been shown that a first order phase transition 
initiated by the nucleation of quark matter droplets is absolutely forbidden 
if the magnetic field $\sim 10^{15}$G at the core region \cite{R9,R10}. However
a second order phase transition is allowed, provided the magnetic field 
strength $<10^{20}$G. This is of course too high to achieve at the core region.

The elementary processes, in particular, the weak and the electromagnetic
processes taking place at the core region of a neutron star are strongly 
affected by such ultra-strong magnetic field \cite{R11,R12}. Since the cooling 
of neutron stars are mainly controlled by neutrino/anti-neutrino emissions, the 
presence of strong quantizing magnetic field should affect the thermal history 
of strongly magnetized neutron stars. Further, the electrical conductivity of 
neutron star matter which directly controls the evolution of neutron star 
magnetic field will also change significantly.

In another kind of work, the stability of such strongly magnetized rotating 
objects are studied. It has been observed from the detailed general 
relativistic calculation that there are possibility of some form of geometrical
deformation of these objects from their usual spherical shapes 
\cite{R13,R14,R15}. In the extreme case such objects may either become 
black strings or black disks. In the non-extreme case, however, it is
also possible to detect gravity waves from these deformed rotating objects.

In a recent study on microscopic model of dense neutron star matter,
we have observed that if most of the electrons occupy
the zeroth Landau level, with spin anti-parallel to the direction of
magnetic field and very few of them are with spin along the direction of
magnetic field and Landau quantum number $>0$, then either such strongly
magnetized system can not exist or such a strong magnetic field is just 
impossible at the core region of a neutron star \cite{R16}. 

Motivated by the problems as mentioned in preceding two sections, in
this paper we shall study the effect of strong quantizing magnetic
field on the transport coefficients of dense stellar electron-proton
plasma. We shall follow an exact formalism \cite{R17} which is applicable for a
wide range of magnetic field strengths and obtain the transport
coefficients from the relativistic version of Boltzmann kinetic equation
by linearizing the distribution function and using relaxation time
approximation. We shall obtain the relaxation time from the rates of
standard electromagnetic processes taking place inside electron-proton
plasma and make necessary modification in the rate calculation due to
the presence of strong quantizing magnetic field.   We have noticed
that the electrical conductivity of the medium becomes 
extremely small in presence of ultra-strong magnetic field ($\geq 10^{17}$G). 
The magnetic field at the core region of a magnetar must therefore decay very 
rapidly (time scale $\sim$ a few mins.) and becomes moderate or low enough. As 
a consequence there will be in principle no problem on the existence of
magnetars (with very low or moderate core magnetic field). The formalism
we have developed to obtain rates of electromagnetic processes or the
relaxation time is also applicable to evaluate neutrino emissivity and
mean free path in presence of strong quantizing magnetic field.

In presence of strong quantizing magnetic field, since the momentum
component in the transverse plane with respect to the external magnetic
field gets quantized, whereas the component along the field direction
varies continuously (from $-\infty$ to $+\infty$), the momentum space volume 
element becomes
\begin{equation}
\frac{d^3p}{(2\pi)^3}=\frac{dp_xdp_ydp_z}{(2\pi)^3}
=\frac{eB_m}{4\pi^2}\sum_{\nu=0}^\infty(2-\delta_{\nu 0}) dp_z
\end{equation}
(we have assumed $\hbar=c=k_b=1$) where we have chosen the gauge $A^\mu\equiv 
(0,0,xB_m,0)$, so that the constant magnetic field $B_m$ is along z-direction.
We have considered the simplest possible picture of neutron star matter with 
$n-p-e$ out of thermodynamic equilibrium and the neutrinos are assumed
to be non-degenerate.  The baryonic components are interacting via
$\sigma-\omega-\rho$ meson exchange type mean field and the
electrons are assumed to be freely moving particles.

In this article, we shall first calculate the transport coefficients of 
electron gas. Then it is very easy to obtain the transport coefficients for 
proton matter just by replacing mass, chemical potential etc. of electrons by 
protons and taking into account the proper modification in presence of 
$\sigma-\omega-\rho$ meson exchange type mean field \cite{R6}. Spinor solutions
for electrons in presence of strong quantizing magnetic fields are then given by
\begin{equation}
\Psi^{(\uparrow)(e)}=\frac{1}{\sqrt {L_yL_z}}
\frac{\exp(-i\varepsilon_{\nu}^{(e)}t+ip_y y+ip_z z)}
{[2\varepsilon_{\nu}^{(e)}(\varepsilon_{\nu}^{(e)}+m_e)]^{1/2}}
\times\left( \begin{array}{c}(\varepsilon_{\nu}^{(e)}+m_e) I_{\nu;p_y}(x)\\
0\\p_z I_{\nu;p_y}(x)\\-i(2\nu e B_m)^{1/2}I_{\nu-1;p_y}(x)
\end{array} \right)
\end{equation}
and
\begin{equation}
\Psi^{(\downarrow)(e)}=\frac{1}{\sqrt {L_yL_z}}
\frac{\exp(-i\varepsilon_{\nu}^{(e)}t+ip_y y+ip_z z)}
{[2\varepsilon_{\nu}^{(e)}(\varepsilon_{\nu}^{(e)}+m_e)]^{1/2}}
\times\left( \begin{array}{c} 0\\(\varepsilon_{\nu}^{(e)}+m_e)I_{\nu-1;p_y}(x)\\
i(2\nu e B_m)^{1/2}I_{\nu;p_y}(x)\\-p_z I_{\nu-1;p_y}(x)
\end{array} \right)
\end{equation}
where $\uparrow$ and
$\downarrow$ represent up and down spin states respectively,
\begin{equation}
I_{\nu;p_y}(x)=\left( \frac{e B_m}{\pi}\right)^{1/4}
\frac{1}{\sqrt{\nu!}~2^{\nu/2}}
\times\exp\left[ -\frac{1}{2}e B_m \left( x- \frac{p_y}{e B_m}\right)^2 
\right]
H_{\nu} \left[ \sqrt{e B_m}\left( x- \frac{p_y}{e B_m} \right) \right],
\end{equation}
with $H_{\nu}$ the Hermite polynomial of order $\nu$ and 
\begin{equation}
\varepsilon_\nu^e=(p_z^2+m_e^2+2e\nu B_m)^{1/2}
\end{equation}
the energy eigen value with
$\nu=0,1,2,.....,$ the Landau quantum numbers.
For neutron we consider the usual spinor solutions. Since the temperature of 
the system is $\ll$ electron chemical potential, we have not considered
the negative energy spinor solutions, In the energy eigen value $p_\perp=
(2\nu eB_m)^{1/2}$ is the transverse component of electron momentum.
To overcome the serious problem on the mechanical stability and hence the 
existence of magnetars, we have studied the variation of transport coefficients,
in particular the electrical conductivity of electron gas / proton
matter in presence of strong quantizing magnetic field and try to show whether 
the electrical conductivity which is solely responsible for the evolution of 
neutron star magnetic field, becomes sufficiently small in presence of ultra 
strong magnetic field.

The paper is organized in the following manner. In section 2, we have
developed the relativistic version of Boltzmann kinetic equation for
fermions in presence of strong quantizing magnetic field and obtain the
expressions for transport coefficients following de Groot \cite{R18}. We shall 
use the relaxation time approximation for the collision term. In section $3$ we
shall evaluate the the relaxation time from the rates of standard
electromagnetic processes taking place at the core region. We shall
incorporate the necessary changes in the rate calculation due to the
presence of strong quantizing magnetic field (or the effect of quantized
Landau levels). In the last section we shall discuss importance of the results.

\section{Boltzmann Equation}
The relativistic version of Boltzmann transport equation is given by
\cite{R18}
\begin{equation}
p_\mu\partial^\mu f+eF^{\mu\nu}p_\nu\frac{\partial f }{\partial p^\mu}+
\Gamma_{\nu\lambda}^\mu p^\nu p^\lambda \frac{\partial f}{\partial p^{\mu}}=C
\end{equation}
where the second and the third terms are coming from electromagnetic and
gravitational (general relativistic) interactions respectively and $C$ is the 
collision term.  Since we have considered the flat space-time geometry and 
noticed that there is no contribution of external magnetic field, only the 
induced electric field arising from local charge non-neutrality, causes an 
induced current in the system will contribute in the electromagnetic 
interaction term and the curvature term is neglected.

To obtain the shear and bulk viscosity coefficients, heat conductivity
and electrical conductivity of dense electron gas we make the relaxation
time approximation, given by
\begin{equation}
C=-\frac{p_0}{\tau}\left (f(x,p)-f^0(p)\right )
\end{equation}
where
\begin{equation}
f(x,p)=f^{(0)}{(p)}\left(1+\chi(x,p)\right)
\end{equation}
i.e., the system is assumed to be very close to the local equilibrium 
configuration. Here $\tau$ is the relaxation time and $f^0$ is
the (local) equilibrium distribution (Fermi distribution) function, given by. 
\begin{equation}
f^{(0)}(p)=\frac{1}{\exp{\beta(\epsilon_\nu-\mu_e)}+1}
\end{equation}

We shall now follow the general technique to obtain the perturbed part
$\chi$ as a linear sum of driving forces. Following the formalism developed in 
the book by de Groot \cite{R18}, we express the four derivative as the sum
of a space like part and a time like part, given by ${\partial}^\mu = u^\mu 
D+{\nabla}^\mu$ with $u^\mu$ the hydrodynamic velocity, $D=u^\mu\partial_\mu$ 
is the convective time derivative and $\nabla^\mu=\Delta^{\mu\nu}\partial_\nu$
is the gradient operator, with $\Delta^{\mu\nu}=g^{\mu\nu}-u^\mu u^\nu$
some kind of projection operator, $g^{\mu\nu}={\rm{diag}}(1, -1, -1, -1)$ the 
metric tensor. Then using the eqns.(6)-(9), the decomposition of $\partial_\mu$
and the equations of motion, given by 
\[
Dn=-n\nabla_\mu u^\mu, ~~ Du^\mu=\frac{1}{nh}\nabla^\mu P, ~~
C_vDT=-F(T)\nabla_\mu u^\mu
\]
we obtain the perturbative part $\chi$, given by
\begin{eqnarray}
\frac{T p^{0}}{\tau}{\chi}&=&QX{\left (1-f^{(0)}\right) }-{\left(p^\mu
u_\mu-h \right )
} p_\nu {X_q}^\nu {\left(1-f^{(0)}\right)}
+ p^\mu p^\nu {X_{\mu\nu}^{0}}{\left(1-f^{(0)}\right)}\nonumber \\
&+&e{\left[{\frac{p_\mu u^\mu}{h}+1}\right]}
p_\nu E^\nu (1- f^{(0)})
\end{eqnarray}
where
\begin{equation}
X=-{\nabla}^\mu u_\mu
\end{equation}
is the driving force for bulk-viscosity,
\begin{equation}
X_q^\mu= \nabla^\mu T-\frac{T}{nh}\nabla^\mu P
\end{equation}
is the driving force for heat conduction,
\begin{equation}
X^{0 \mu\nu}= \nabla^\mu u^\nu-\frac{1}{3}\Delta^{\mu \nu}\nabla_\sigma u^\sigma
\end{equation}
is the driving force for shear viscosity and $E^\nu$ is the driving force for
electric current ($E^\nu$ for $\nu=i=1,2,3$ are the components of electric field
vector). The quantity $Q$ is given by
\begin{equation}
Q=-\frac{1}{3}\Delta^{\mu\nu}p_\mu p_\nu+(p^\mu u_\mu)^2 
\frac{F(T)}{T}(1-\gamma ) 
+\left(T^2(1-\gamma)\frac{F(T)}{T}\frac{\partial}{\partial T}
\left(\frac{\mu}{T}\right)-n \frac{\partial\mu}{\partial n}\right)
p^\mu u_\mu
\end{equation}
with $F(T)=P(T)/n(T)$, $P(T)$ and $n(T)$ are the
equilibrium local kinetic pressure and number density respectively. For a
non-relativistic Boltzmann gas $F(T)=T$, the local temperature of the
system and finally $h=(\epsilon +P)/n$, the enthalpy per particle with
$\epsilon$ the local energy density and $\gamma=C_p/C_v$
the ratio of specific heats at constant pressure and constant volume
respectively.

Now from the definition, the heat flow four current is given by
\begin{eqnarray}
I_q^\mu&=&\frac{eB_m}{4\pi^2}\sum_\nu(2-\delta_{\nu 0})
\int_{-\infty}^{+\infty} 
\frac{dp_z}{p_0}\left(p^\nu u_\nu-h\right)
p^\mu f(x,p) \nonumber\\
&=&I_q^{(0)\mu}+I_q^{(1)\mu}
\end{eqnarray}
where the first term is the equilibrium contribution, which is identically 
zero. Then omitting the symbol $(1)$, we have the irreversible term
\begin{equation}
I_q^{(1)\mu} = I_q^\mu 
= \frac{eB_m}{4\pi^2}\sum_\nu (2-\delta_{nu 0})\int_{-\infty}^{+\infty}\
\frac{dp_z}{p_0}p^\mu{\left(p^\nu u_\nu-h\right)}
f^{(0)}(p)\chi(x,p)
\end{equation}
Again using the definition
\begin{equation}
{I_q}^\mu=\lambda^{\mu\nu} X_{q\nu},
\end{equation}
we have the heat conductivity coefficient
\begin{equation}
\lambda = \frac{1}{3} \Delta^{\mu\nu}\lambda_{\mu\nu}
= -\frac{1}{3T^2}\frac{eB_m}{4\pi^2}\sum_\nu (2-\delta_{\nu 0})
\int_{-\infty}^{+\infty} \frac{dp_z}{p_0^2}\tau
(p^\sigma u_\sigma-h)^2 f^{(0)}(1- f^{(0)})
\Delta^{\mu\nu} p_\mu p_\nu
\end{equation}
Now from the definition, the energy-momentum tensor is given by
\begin{equation}
T^{\mu\nu}=\frac{eB_m}{4\pi^2}\sum_\nu (2-\delta_{\nu
0})\int_{-\infty}^{+\infty}
\frac{dp_z}{p_0}p^\mu p^\nu f(x,p)
\end{equation}
Which can also be written as $T^{\mu\nu}=T^{(0)\mu\nu}+ T^{(1)\mu\nu}$, with
the equilibrium value
\begin{equation}
T^{(0)\mu\nu}=\frac{eB_m}{4\pi^2}\sum_\nu (2-\delta_{\nu 0})
\int_{-\infty}^{+\infty}\frac{dp_z}{p_0}p^\mu p^\nu
f^{(0)}(p)
\end{equation}
and the non-equilibrium part
\begin{equation}
T^{(1)\mu\nu}=T^{\mu\nu}=\frac{eB_m}{4\pi^2}\sum_\nu (2-\delta_{\nu 0})
\int_{-\infty}^{+\infty}\frac{dp_z}{p_0}p^\mu p^\nu
\chi(x,p)f^{(0)}(p)
\end{equation}
which is a symmetric second rank tensor. Now we consider a model in which a 
flow of electron gas with cylindrical symmetry is assumed. Then considering
$\mu=r$, $\nu=z$, we have  from the definition
\begin{equation}
T^{(1)rz}=-\eta\frac{du_z}{dr}
\end{equation}
Hence the shear viscosity coefficient is given by
\begin{equation}
\eta=\frac{eB_m}{4\pi^2T}\sum_\nu (2-\delta_{\nu 0})
\int_{-\infty}^{+\infty}\frac{dp_z}{p_0^2}(p^r p^z)^2 
f^{(0)}(1- f^{(0)})\tau
\end{equation}
where $p_r=(2\nu eB_m)^{1/2}$, the transverse component of electron momentum. 
Then it is quite obvious that in presence of ultra strong magnetic field, for 
which $\nu_{\rm{max}}=0$, the shear viscosity coefficient vanishes. Here 
$\nu_{\rm{max}}$ is the maximum value of quantum number of the Landau levels 
occupied by electrons for a given density and temperature. For $T=0$, 
\[
\nu_{\rm{max}}=\left [ \frac{\mu_e^2-m_e^2}{2eB_m} \right ],
\]
where $[~]$ indicates the nearest integer less than the actual value.

To obtain an expression for bulk viscosity coefficient, we next consider the 
pressure tensor, given by
\begin{equation}
{\Pi}^{\mu\nu}=\Delta_\sigma^\mu T^{\sigma\tau}\Delta_\tau^\nu+
P\Delta^{\mu\nu}
\end{equation}
Where the reversible part
\begin{equation}
\Pi^{(0) \mu\nu}=0
\end{equation}
and the non-equilibrium part
\begin{eqnarray}
\Pi^{(1)\mu\nu}&=&\Delta_\sigma^\mu T^{\sigma\tau (1)}\Delta_\tau^\nu
\nonumber \\
&=&\frac{eB_m}{4\pi^2}\sum_\nu (2-\delta_{\nu 0})
\int_{-\infty}^{+\infty}\frac{dp_z}{p_0}\Delta_\sigma^\mu \Delta_\tau^\nu
p^\sigma p^\tau \chi(x,p)f^{(0)}(p) 
\end{eqnarray}
Hence omitting $1$, we have the traceless part of pressure tensor
\begin{eqnarray}
\Pi &=& -\frac{1}{3}\Pi_\mu^\mu\nonumber \\
&=&-\frac{1}{3}\frac{eB_m}{4\pi^2}\sum_\nu(2-\delta_{\nu 0})
\int_{-\infty}^{+\infty} \frac{dp_z}{p_0}\Delta_{\sigma \mu} 
\Delta_\tau^\mu p^\sigma p^\tau
f^{(0)}(p)\chi(x,p)\nonumber \\
&=&\eta_v\nabla_\mu u^\mu
\end{eqnarray}
Hence we have the bulk viscosity coefficient
\begin{equation}
\eta_v =\frac{1}{3T}\frac{eB_m}{4\pi^2}\sum_\nu
(2-\delta_{\nu 0})
\int_{-\infty}^{+\infty}\frac{dp_z}{p_0^2}\Delta_{\tau\sigma} 
p^\tau p^\sigma \tau
Q (1- f^{(0)})f^{(0)}
\end{equation}

We shall now calculate the the electrical conductivity for electron gas.
The electric four current is given by
\begin{equation}
j^\mu(x)=\frac{eB_m}{4\pi^2}\sum_\nu (2-\delta_{\nu 0})
\int_{-\infty}^{+\infty}\frac{dp_z}{p_0} p^\mu f(x,p)
\end{equation}
Now because of local charge neutrality, the reversible part
$j^{\mu (0)}=0$
and the non-equilibrium part is given by
\begin{equation}
j^{\mu (1)}=j^\mu=\frac{eB_m}{4\pi^2}\sum_\nu (2-\delta_{\nu 0})
\int_{-\infty}^{+\infty}\frac{dp_z}{p_0} p^\mu f^{(0)} \chi(x,p)
\end{equation}
Then using the covariant form of Ohm's law
\begin{equation}
j^\mu=\sigma^{\mu\nu} E_\nu
\end{equation}
the electrical conductivity tensor is given by
\begin{equation}
\sigma^{\mu\nu}=\frac{e^2}{T}\frac{eB_m}{4\pi^2}\sum_\nu (2-\delta_{\nu 0})
\int_{-\infty}^{+\infty}\tau\frac{dp_z}{p_0^2}p^\mu p^\nu 
\left(\frac{p^\alpha u_\alpha}{h}+1\right) f^{(0)}(1- f^{(0)}) 
\end{equation}   
This equation shows that due the presence of strong quantizing magnetic
field the electrical conductivity becomes a second rank tensor- even if
the space is isotropic in nature. The $zz$-component is given by
\begin{equation}
\sigma^{zz}=\frac{e^2}{T}\frac{eB_m}{4\pi^2}\sum_\nu (2-\delta_{\nu 0})
\int_{-\infty}^{+\infty}{dp_z}\frac{\tau}{p_0^2} p_z^2
\left(1+\frac{p^\alpha u_\alpha}{h}\right)
f^{(0)}(1- f^{(0)})
\end{equation}
The $\perp\perp$-component is given by
\begin{equation}
\sigma^{\bot\bot}=\frac{e^2}{T}\frac{eB_m}{4\pi^2}\sum_\nu(2-\delta_{\nu
0})
(2\nu eB_m)\int_{-\infty}^{+\infty}{dp_z}\frac{\tau}{p_0^2}
\left(1+\frac{p^\alpha u_\alpha}{h}\right)
f^{(0)}(1- f^{(0)}) 
\end{equation}
and finally the $\perp z$-component is given by
\begin{equation}
\sigma^{\bot z}=\sigma^{z \bot}
=\frac{e^2}{T}\frac{eB_m}{4\pi^2}\sum_\nu(2-\delta_{\nu 0})
(2\nu eB_m)^{1/2}\int_{-\infty}^{+\infty}{dp_z}\frac{\tau p_z}{p_0^2}
\left(1+\frac{p^\alpha u_\alpha}{h}\right) f^{(0)}(1- f^{(0)})
\end{equation}
From eqn.(34) and (35) it is quite obvious that just like the shear
viscosity coefficient, both $\sigma^{\perp\perp}$ and $\sigma^{\perp z}$
components of electrical conductivity vanish for $\nu_{\rm{max}}=0$, i.e, in
the ultra strong magnetic field limit. This is of course not at all
evident for $\eta_v$, $\lambda$ and $\sigma^{zz}$.

\section{Rate of Electromagnetic Processes}
Now to obtain the numerical values of all these transport coefficients,
or their variations with the strength of magnetic field we have to know 
the relaxation time $\tau$, given by
\begin{equation}
\frac{1}{\tau}=\sum_i W_i
\end{equation}
where $W_i$ is the rate of $i^{th}$ electromagnetic process and the sum 
is over all possible electromagnetic processes taking
place involving the electrons. 

Therefore to obtain the relaxation time we evaluate the rates of the
basic electromagnetic processes, given by $e+e \rightarrow e+e,~{\rm{and}}~
e+p\rightarrow e+p$.  Now in the case of $e-e$-Scattering it is necessary to 
consider both direct and exchange processes, whereas in the case of $e-p$ 
scattering only the direct term contributes. Let us first consider the 
$e-e$-scattering process, then the direct part is given by
\begin{eqnarray}
T_{\rm{fi}}^{(d)}& =&
\frac{ie^2}{L_y^2L_z^2Q^2}(2\pi)^3\delta(E_1+E_2-E_3-E_4)
\delta(k_{1y}+k_{2y}-k_{3y}-k_{4y})
\delta(k_{1z}+k_{2z}-k_{3z}-k_{4z})\nonumber \\
&&\int_{-\infty}^\infty dx \left(\bar u(k_3,x)\gamma_\mu
u(k_1,x)\right)\left(\bar
u(k_4,x)\gamma^\mu u(k_2,x)\right)
\end{eqnarray}
Similarly the exchange term is given by
\begin{eqnarray}
T_{\rm{fi}}^{(ex)} &=&
\frac{-ie^2}{L_y^2L_z^2Q^2}(2\pi)^3\delta(E_1+E_2-E_3-E_4)
\delta(k_{1y}+k_{2y}-k_{3y}-k_{4y})
\delta(k_{1z}+k_{2z}-k_{3z}-k_{4z})\nonumber \\
&&\int_{-\infty}^\infty dx \left(\bar u(k_4,x)\gamma_\mu
u(k_1,x)\right)\left(\bar
u(k_3,x)\gamma^\mu u(k_2,x)\right)
\end{eqnarray}
where $Q$ is the exchanged momentum. The rate element is then given by
\begin{equation}
dW=\lim_{t\rightarrow \infty}\frac{\mid T_{\rm{fi}}^{(d)}+
T_{\rm{fi}}^{(ex)} \mid^2}{t}
\end{equation}
Since the spinors are functions of $x$-coordinate through
$I_{\nu,p_y}(x)$ (see eqns.(2)-(4)), we use the relation
\begin{equation}
\mid f \mid^2=\int_{-\infty}^\infty dx\int_{-\infty}^\infty dx^\prime
f^*(x)f(x^\prime)
\end{equation}
and we also need the positive energy projection operator, given by
\begin{equation}
\Lambda^+=\sum_{\rm{spin}}u(k,x)\bar u(k,x^\prime)
\end{equation}
Substituting the positive energy spinor solutions (eqns.(2)-(3)), we have
\begin{equation}
\Lambda^+=\frac{1}{2E_\nu}(Ak_\mu\gamma^\mu (\mu=0 ~~{\rm{and}}~~
z)\hfil\break+mA+B k_\mu\gamma^\mu (\mu=y ~~{\rm{and}}~~ p_y=p_\perp)
\end{equation}
The matrices $A$ and $B$ are given by
\begin{equation}
A=\left ( \begin{array}{l c c r}I_\nu I_\nu^\prime &0&0&0 \\
0 & I_{\nu-1}I_{\nu-1}^\prime &0 &0 \\
0 & 0 &I_\nu I_\nu^\prime &0 \\
0 & 0 & 0 &I_{\nu-1}I_{\nu-1}^\prime \\
\end{array}
\right )
\end{equation}
\begin{equation}
B= \left ( \begin{array}{l c c r}I_{\nu-1} I_\nu^\prime &0&0&0 \\
0 & I_\nu I_{\nu-1}^\prime &0 &0 \\
0 & 0 &I_{\nu-1} I_\nu^\prime &0 \\
0 & 0 & 0 &I_\nu I_{\nu-1}^\prime \\
\end{array}
\right )
\end{equation}
where primes indicate the functions of $x'$. Since the Dirac $\gamma$ matrices 
are traceless and both $A$ and $B$ matrices are diagonal in nature with 
identical blocks we have a few interesting relations, e.g., 
\begin{equation}
{\rm{Tr}}(\gamma^\mu \gamma^\nu A_1A_2..B_1B_2..)=Tr(A_1A_2..B_1B_2..)
g^{\mu\nu},
\end{equation}
\begin{equation}
{\rm{Tr}}(\gamma^\mu\gamma^\nu\gamma^\lambda\gamma^\sigma
A_1A_2..B_1B_2..)=Tr(A_1A_2..B_1B_2..)(g^{\mu \nu}
g^{\sigma \lambda}-g^{\mu \lambda}g^{\nu \sigma}+g^{\mu \lambda}g^{\nu
\sigma}) 
\end{equation}
\begin{equation}
{\rm{Tr}}({\rm{product~ of ~odd~ no ~of}}\gamma{\rm{s ~with~  any~
number~ of~ }}A{\rm{~ and/or~}} B {\rm~{matrices}})=0 
\end{equation}
etc.  The beautiful form of $A$ and $B$ matrices allow to multiply 
$\gamma$-matrices in the above relations from any side or in any order. The 
other interesting aspects of $A$ and $B$ matrices are\\
i) $k_{1\mu}k^{2\mu}{\rm{Tr}}(A_1A_2)= (E_1E_2-k_{1z}k_{2z}){\rm{Tr}}(A_1A_2)$\\
ii) $k_{1\mu}k^{2\mu}{\rm{Tr}}(B_1B_2)= \vec k_{1\perp}.\vec
k_{2\perp}{\rm{Tr}}(B_1B_2)$\\
iii) $k_{1\mu}k^{2\mu}{\rm{Tr}}(A_1B_2)= k_{1\mu}k^{2\mu}{\rm{Tr}}(B_1A_2)=0$\\
iv) $p_{1\mu}k^{1\mu}p_{2\nu}k^{2\nu}{\rm{Tr}}(A_1B_2)\neq 0=
(E_{\nu_1}E_{\nu_2^\prime}-p_{1z}k_{1z})\vec p_{2\perp}.\vec k_{2\perp}
{\rm{Tr}}(A_1B_2)$\\
All these results are entirely new and to our knowledge not reported before 
in the literature. We do believe that these new results can have interesting 
applications in various studies of properties of strongly magnetized dense 
stellar matter. With all these new formulae, the transition matrix element for 
$e~e$ scattering direct term is given by
\begin{eqnarray}
W_{\rm{ee;d}}&=&\sum_{\nu_2}\sum_{\nu_3}\sum_{\nu_4}\int dx dx^\prime
\frac{e^4}{(2\pi)^3}\frac{1}{4}\delta(E_{\nu_1} +E_{\nu_2} -E_{\nu_3}
-E_{\nu_4})\delta(p_{1y}+p_{2y}
-p_{3y}-p_{4y})\delta(p_{1z}+p_{2z}-p_{3z}-p_{4z})\nonumber \\ &&
\gamma_{\nu_2}\gamma_{\nu_3}\gamma_{\nu_4}\frac{1}{(p_1-p_3)^4}\times 
\frac{1}{8E_{\nu_1}E_{\nu_2}E_{\nu_3}E_{\nu_4}}
[(k_3.k_4)(k_1.k_2)\nonumber \\&& \{Tr(A_3A_1)Tr(A_4A_2)
+ Tr(A_3B_1)Tr(A_4B_2) +Tr(B_3A_1)Tr(B_4A_2) +Tr(B_3B_1)Tr(B_4B_2)\}\nonumber \\
&&+(k_3.k_2)(k_1.k_4)\{Tr(A_3A_1)Tr(A_4A_2) + Tr(A_3B_1)Tr(B_4A_2)
+Tr(B_3A_1)Tr(A_4B_2) +Tr(B_3B_1)Tr(B_4B_2)\}\nonumber \\
&&-m_e^2(k_3.k_1)\{Tr(A_3A_1)Tr(A_4A_2) +Tr(B_3B_1)Tr(A_4A_2)\} 
-m_e^2(k_4.k_2)\{Tr(A_3A_1)Tr(B_4B_2)\nonumber \\
&&+Tr(A_3A_1)Tr(A_4A_2)\} +2m_e^4\{Tr(A_3A_1)Tr(A_4A_2)\} ] 
dp_{2y} dp_{2z} dp_{3y} dp_{3z} dp_{4y} dp_{4z}\nonumber \\ &&f_0(p_{2z})
(1-f_0(p_{3z})) (1-f_0(p_{4z}))
\end{eqnarray}
where $\gamma_{\nu_i}={(2-\delta_{\nu_i}0)}$.

The exchange term is given by
\begin{eqnarray}
W_{\rm{ee;ex}}&=&\sum_{\nu_2}\sum_{\nu_3}\sum_{\nu_4}\int dx dx^\prime
\frac{e^4}{(2\pi)^3}\frac{1}{4}\delta(E_{\nu_1} +E_{\nu_2} -E_{\nu_3}
-E_{\nu_4})\delta(p_{1y}+p_{2y}
-p_{3y}-p_{4y})\delta(p_{1z}+p_{2z}-p_{3z}-p_{4z})\nonumber \\ &&
\gamma_{\nu_2}\gamma_{\nu_3}\gamma_{\nu_4}\frac{1}{(p_1-p_4)^4}
\times \frac{1}{8E_{\nu_1}E_{\nu_2}E_{\nu_3}E_{\nu_4}}
[(k_3.k_4)(k_1.k_2)\nonumber \\&& 
\{Tr(A_1A_4)Tr(A_2A_3)+Tr(B_1A_4)Tr(B_2A_3)+Tr(A_1B_4)Tr(A_2B_3)+Tr(B_1B_4)
Tr(B_2B_3)\}\nonumber \\
&+&(k_1.k_3)(k_2.k_4)\{Tr(A_1A_4)Tr(A_2A_3)+Tr(B_1A_4)Tr(A_2B_3)
+Tr(A_1B_4)Tr(B_2A_3)+Tr(B_1B_4)Tr(B_2B_3)\}\nonumber \\
&-&m_e^2(k_1.k_4)\{Tr(A_1A_4)Tr(A_2A_3)+Tr(B_1B_4)Tr(A_2A_3)\} \nonumber
\\ &-&m_e^2(k_2.k_3)\{Tr(A_1A_4)Tr(A_2A_3)+Tr(A_1A_4)Tr(B_2B_3)\}\nonumber
\\ &+&2m_e^4Tr(A_1A_4)Tr(A_2A_3)]
dp_{2y} dp_{2z} dp_{3y} dp_{3z} dp_{4y} dp_{4z}\nonumber \\
&&f_0(p_{2z}) (1-f_0(p_{3z})) (1-f_0(p_{4z}))
\end{eqnarray}
and similarly the $e~e$-scattering mixed term is given by
\begin{eqnarray}
W_{\rm{ee;mix}}&=&\sum_{\nu_2}\sum_{\nu_3}\sum_{\nu_4}\int dx dx^\prime
\frac{e^4}{(2\pi)^3}\frac{1}{4}\delta(E_{\nu_1}+ E_{\nu_2}-E_{\nu_3}-
E_{\nu_4})\delta(p_{1y}+p_{2y}
-p_{3y}-p_{4y})\delta(p_{1z}+p_{2z}-p_{3z}-p_{4z})\nonumber \\ &&
\gamma_{\nu_2}\gamma_{\nu_3}\gamma_{\nu_4}\frac{1}{(p_1-p_4)^2(p_1-p_3)^2}
\times \frac{1}{8E_{\nu_1} E_{\nu_2}E_{\nu_3}E_{\nu_4}}
[-4(k_3.k_4)(k_1.k_2)\nonumber \\&& \{Tr(A_1A_2A_3A_4)+Tr(A_1A_2B_3B_4)+
Tr(B_1B_2A_3A_4)+Tr(B_1B_2B_3B_4)\} \nonumber \\ 
&+& 2m_e^2(k_1.k_3)\{Tr(A_1A_2A_3A_4)+Tr(B_1A_2B_3A_4)
+2m_e^2(k_1.k_2)\{Tr(A_1A_2A_3A_4)+Tr(B_1B_2A_3A_4)\} ] \nonumber \\
&+& 2m_e^2(k_1.k_4)\{Tr(A_1A_2A_3A_4)+Tr(B_1A_2A_3B_4)\}
+2m_e^2(k_2.k_3)\{Tr(A_1A_2A_3A_4)+Tr(A_1B_2B_3B_4)\}\nonumber \\
&+& 2m_e^2(k_3.k_4)\{Tr(A_1A_2A_3A_4)+Tr(A_1A_2B_3B_4)\}+\nonumber \\
&&2m_e^2(k_2.k_4)\{Tr(A_1A_2A_3A_4)Tr(A_1B_2A_3B_4)\}-m_e^4Tr(A_1A_2A_3A_4) ]
\nonumber \\ &&dp_{2y} dp_{2z} dp_{3y} dp_{3z} dp_{4y} dp_{4z}
f_0(p_{2z}) (1-f_0(p_{3z})) (1-f_0(p_{4z}))
\end{eqnarray}
and finally the rate of $e~p$-scattering is given by
\begin{eqnarray}
W_{\rm{ep}}&=&\sum_{\nu_2}\sum_{\nu_3}\sum_{\nu_4}\int dx dx^\prime
\frac{e^4}{(2\pi)^3}\frac{1}{4}\delta(E_1+E_2-E_3-E_4)\delta(p_{1y}+p_{2y}
-p_{3y}-p_{4y})\delta(p_{1z}+p_{2z}-p_{3z}-p_{4z})\nonumber \\ &&
\gamma_{\nu_2}\gamma_{\nu_3}\gamma_{\nu_4}\frac{1}{(p_1-p_3)^4}
\times \frac{1}{8E_{\nu_1}E_{\nu_2}E_{\nu_3}E_{\nu_4}}
[(k_3.k_4)(k_1.k_2)\nonumber \\&& \{Tr(A_3A_1)Tr(A_4A_2)
+ Tr(A_3B_1)Tr(A_4B_2) +Tr(B_3A_1)Tr(B_4A_2) +Tr(B_3B_1)Tr(B_4B_2)\}\nonumber \\
&&+(k_3.k_2)(k_1.k_4)\{Tr(A_3A_1)Tr(A_4A_2) + Tr(A_3B_1)Tr(B_4A_2)
+Tr(B_3A_1)Tr(A_4B_2)\nonumber \\ &&+Tr(B_3B_1)Tr(B_4B_2)\} -m_p^2(k_3.k_1)
\{Tr(A_3A_1)Tr(A_4A_2) +Tr(B_3B_1)Tr(A_4A_2)\} \nonumber \\ &&-m_e^2(k_4.k_2)
\{Tr(A_3A_1)Tr(B_4B_2) +Tr(A_3A_1)Tr(A_4A_2)\} + 2m_e^2m_p^2\{Tr(A_3A_1)
Tr(A_4A_2)\} ]\nonumber \\ && dp_{2y} dp_{2z} dp_{3y} dp_{3z} dp_{4y} dp_{4z}
f_0(p_{2z}) (1-f_0(p_{3z})) (1-f_0(p_{4z}))
\end{eqnarray}

To evaluate the multidimensional (eight dimensional) integrals, we have
used three $\delta$-functions and finally evaluate numerically using
multi-dimensional Monte-Carlo integration technique and obtain the rates from
the eqns.(48)-(51). We have generated the Hermite polynomials
appearing in the matrices A and B numerically.  We obtain the relaxation time 
from eqn.(36) and finally evaluate the kinetic coefficients from 
eqns.(18),(23),(28),(33)-(35) for various values of magnetic fields $B_m$, 
temperature $T$ and matter density $n_B$. In figs.(1)-(6) we have shown the 
variation of kinetic coefficients with magnetic field strength. Identical 
results can also be obtained for dense proton matter.

\section{Discussions}
From the figs.(1)-(6) it is quite obvious that the kinetic coefficients 
are almost independent of magnetic field for moderate strengths
$(<10^{17} G)$,  but all of them go to almost zero for magnetic field
strength beyond $10^{17}G$. The first conclusion is therefore, that at
ultra strong magnetic field, the matter (electron or proton matter)
behaves like a superfluid insulator. The mechanism of superfluidity is
of course completely different from conventional neutron matter.

Now as we know from classical plasma physics in presence of strong
magnetic field, that the charged particles can only travel along the
lines of force, motions are almost forbidden across the field, as a
consequence $\sigma_{\perp\perp}$ or $\sigma_{\perp z}$ vanish for a
classical plasma in presence of ultra-strong magnetic field, whereas
$\sigma_{zz}$ remains non-zero even at very high magnetic field
strength. On the other hand, in the case of quantum mechanical plasma as
the magnetic field become strong enough, the maximum value of Landau
quantum number $\nu_{\rm{max}}\rightarrow 0$, which further means the system
becomes effectively one dimensional in the ultra-strong magnetic field
limit. Then in the collision dominated scenario, since $p_z$ varies
continuously from $-\infty$ to $+\infty$ and $p_\perp\rightarrow 0$,
$\sigma_{zz}\rightarrow 0$. To elaborate this point a bit- since the current is
flowing along the field lines only, then from the symmetric nature of
$p_z$, we have
\begin{equation}
j_z^{(+)}=-j_z^{(-)}\Longrightarrow j_z=j_z^{(+)}+j_z^{(-)}=0
\end{equation}
This is specially true for an effectively one dimensional collision
dominated quantum plasma. In the case one dimensional flow (free streaming) of 
charged particles along the field lines, this is not true; $\sigma_{zz}$ has
non-zero finite value.  Since all the three component of electrical conductivity
vanish in presence of ultra-relativistic magnetic field, this must affect
significantly the Ohmic decay of strong magnetic field. We know that the
Ohmic decay time scale is given by
\begin{equation}
\tau_d=\frac{4\pi\sigma l^2}{c^2}
\end{equation}
where $l$ is the dimension of the system. Since $\sigma\rightarrow 0$ for 
extremely strong magnetic field, as a consequence, the field should decay 
quickly. We have noticed that a field of strength $10^{18}$G become 
$\sim 10^{15}$G within a minute. This moderate field of course will take 
longer time to decay further (long decay time scale). Therefore the last 
conclusion is that stronger the magnetic field, shorter the decay time scale 
and as a result we do believe that the magnetic field strength at the core 
region of a magnetar can not be high enough and consequently there will be no
problem on the mechanical stability or the existence of magnetars.
\begin{figure}
\psfig{figure=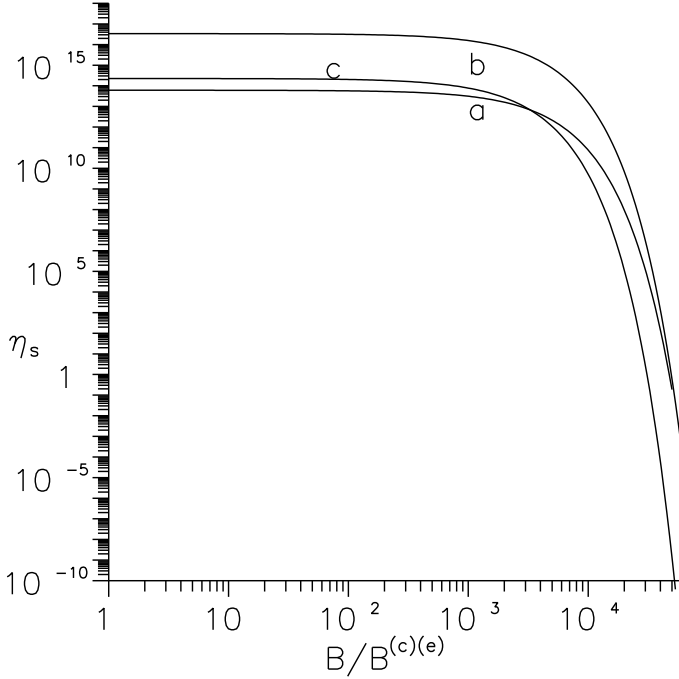,height=0.5\linewidth}
\caption{
Variation of shear viscosity coefficient with magnetic field $B$.
Curve (a): $T=15$MeV and $n_B=3n_0$, Curve (b): $T=15$MeV and $n_B=5n_0$,
and Curve (c): $T=30$MeV and $n_B=3n_0$,
}
\end{figure}
\begin{figure}
\psfig{figure=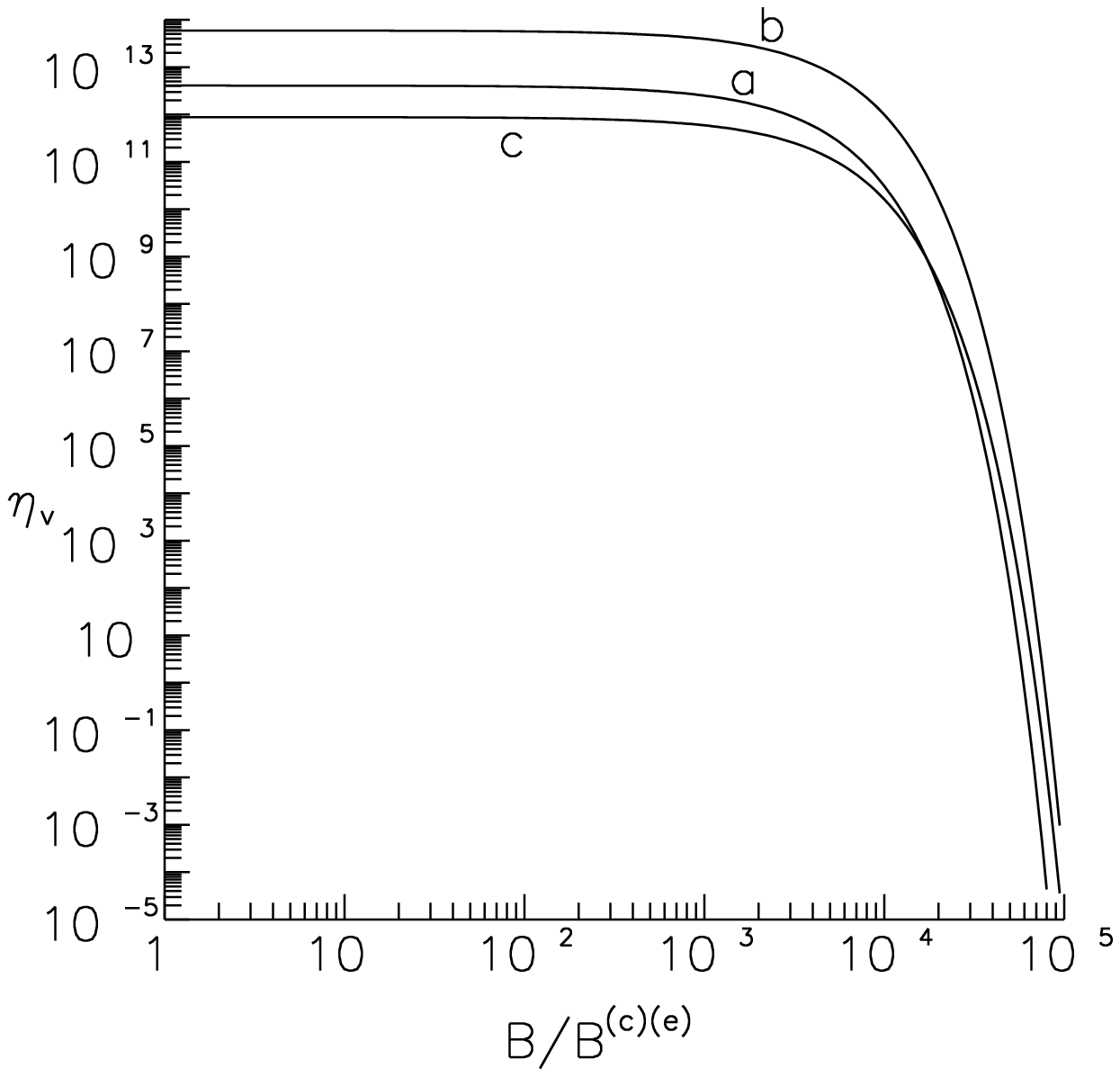,height=0.5\linewidth}
\caption{
Variation of bulk viscosity coefficient with magnetic field $B$.
Curve (a): $T=15$MeV and $n_B=3n_0$, Curve (b): $T=15$MeV and $n_B=5n_0$,
and Curve (c): $T=30$MeV and $n_B=3n_0$,
}
\end{figure}
\begin{figure}
\psfig{figure=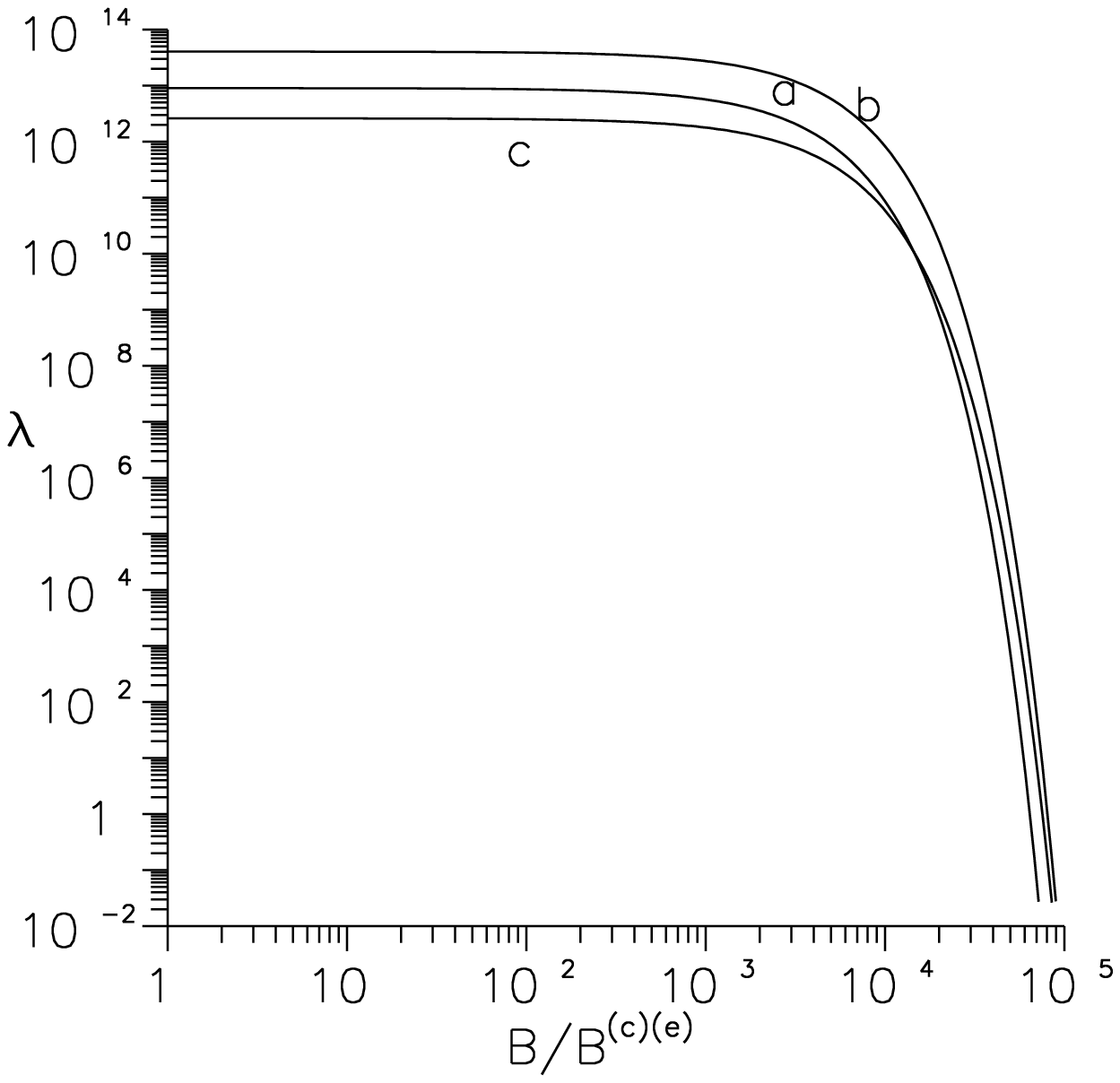,height=0.5\linewidth}
\caption{
Variation of heat conductivity coefficient  with magnetic field $B$.
Curve (a): $T=15$MeV and $n_B=3n_0$, Curve (b): $T=15$MeV and $n_B=5n_0$,
and Curve (c): $T=30$MeV and $n_B=3n_0$,
}
\end{figure}
\begin{figure}
\psfig{figure=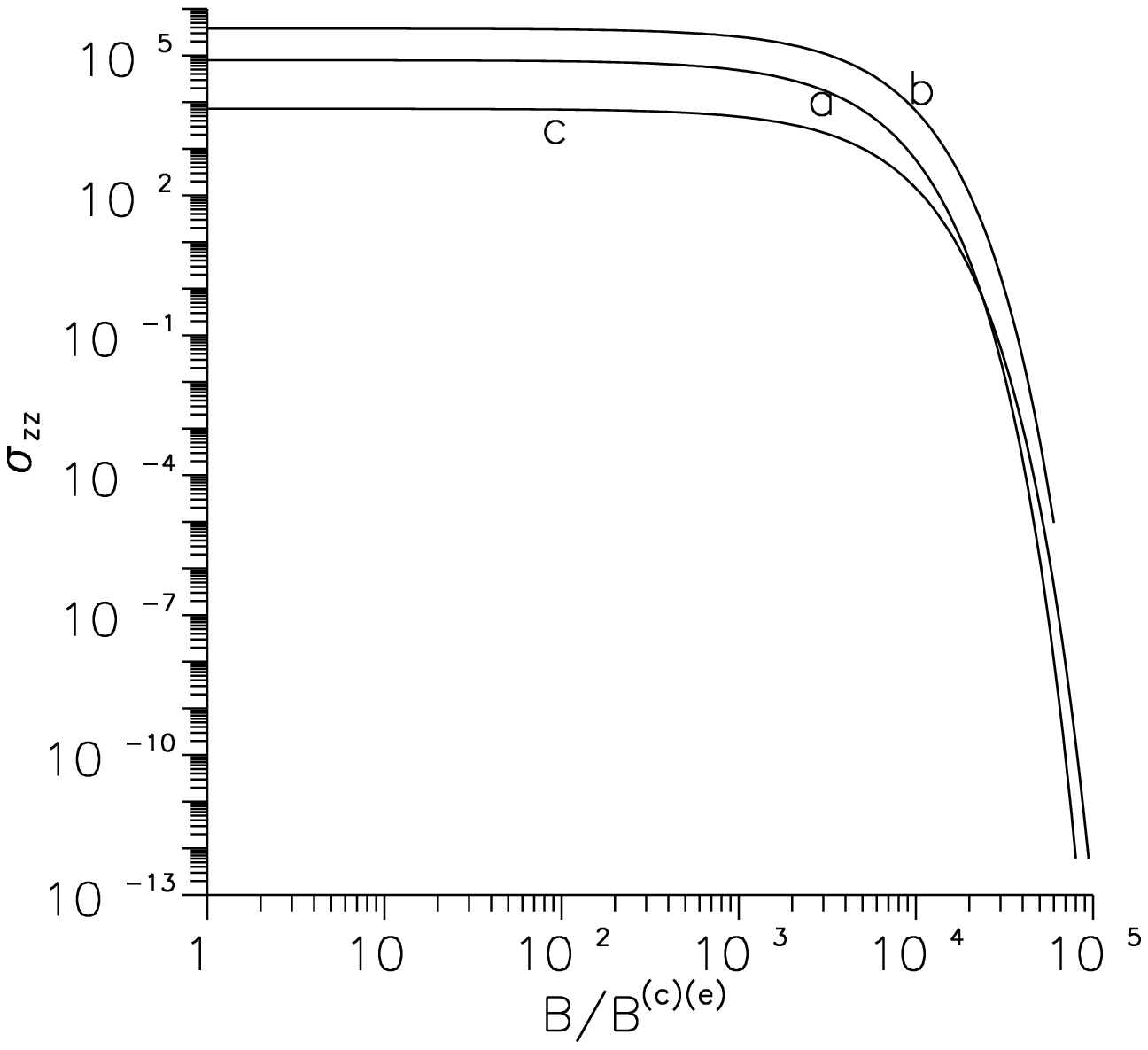,height=0.5\linewidth}
\caption{
Variation of {\bf{zz}} component of electrical conductivity  with magnetic 
field $B$. Curve (a): $T=15$MeV and $n_B=3n_0$, Curve (b): $T=15$MeV and 
$n_B=5n_0$, and Curve (c): $T=30$MeV and $n_B=3n_0$,
}
\end{figure}
\begin{figure}
\psfig{figure=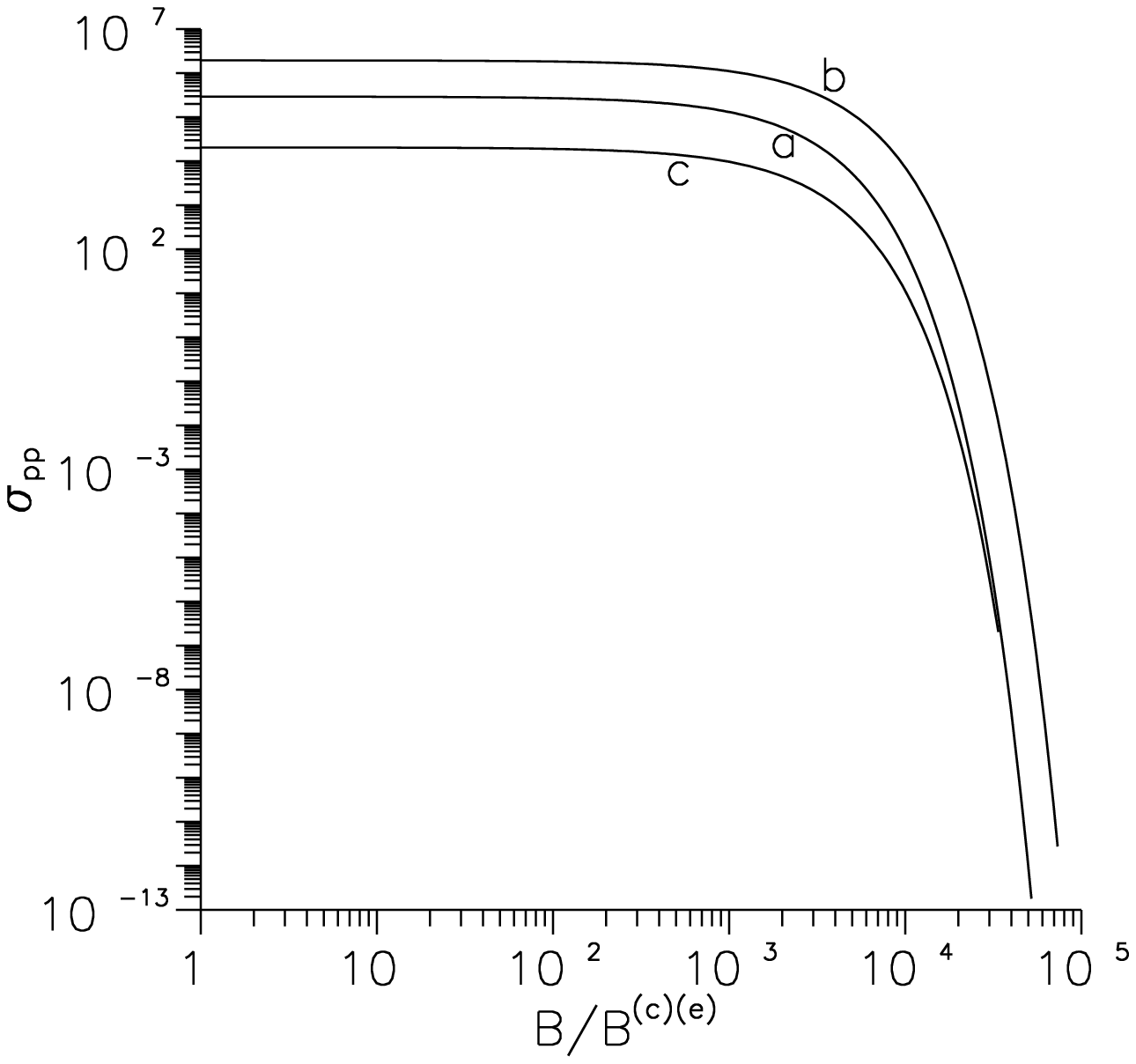,height=0.5\linewidth}
\caption{
Variation of {\bf{pp}} component of electrical conductivity  with magnetic 
field $B$. Curve (a): $T=15$MeV and $n_B=3n_0$, Curve (b): $T=15$MeV and 
$n_B=5n_0$, and Curve (c): $T=30$MeV and $n_B=3n_0$,
}
\end{figure}
\begin{figure}
\psfig{figure=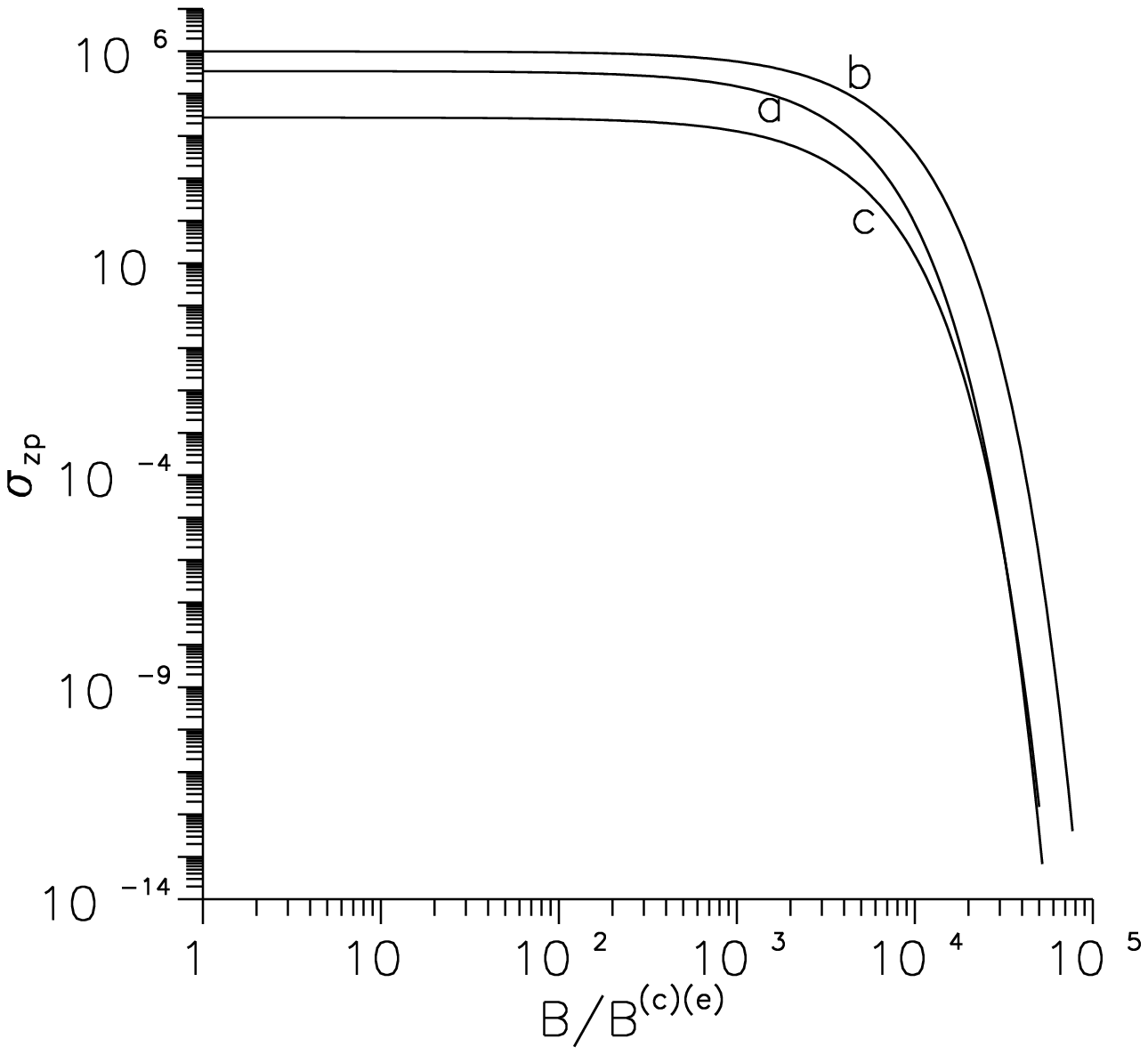,height=0.5\linewidth}
\caption{
Variation of {\bf{zp}} component of electrical conductivity  with magnetic 
field $B$. Curve (a): $T=15$MeV and $n_B=3n_0$, Curve (b): $T=15$MeV and 
$n_B=5n_0$, and Curve (c): $T=30$MeV and $n_B=3n_0$,
}
\end{figure}

\end{document}